# Mapping the youngest and most massive stars in the Tarantula nebula with MUSE-NFM

N. Castro[1]
M. M. Roth[1]
P. M. Weilbacher[1]
G. Micheva[1]
A. Monreal-Ibero[2,3]
A. Kelz[1]
S. Kamann[4]
M. V. Maseda[5]
M. Wendt[6]
and the MUSE collaboration

[1] Leibniz-Institut für Astrophysik Potsdam (AIP), An der Sternwarte 16, 14482 Potsdam, Germany
[2] Instituto de Astrofísica de Canarias (IAC), E-38205 La Laguna, Tenerife, Spain
[3] Universidad de La Laguna, Dpto. Astrofísica, E-38206 La Laguna, Tenerife, Spain
[4] *Astrophysics Research Institute, Liverpool John Moores University, 146 Brownlow Hill, Liverpool, L3 5RF, UK*
[5] Leiden Observatory, Leiden University, P.O. Box 9513, 2300 RA, Leiden, The Netherlands
[6] Institut für Physik und Astronomie, Universität Potsdam, Karl-Liebknecht-Str. 24/25, 14476 Golm, Germany

**The evolution of the most massive stars is a puzzle with many missing pieces. Statistical analyses are the key to provide anchors to calibrate theory, however performing these studies is an arduous job. The state-of-the-art integral field spectrograph MUSE has stirred up stellar astrophysicists who are excited about the capability to take spectra of up to a thousand stars in a single exposure. The excitement was even higher with the commissioning of the MUSE narrow-field-mode (NFM) that has demonstrated angular resolutions akin to the Hubble Space Telescope. We present the first mapping of the dense stellar core R136 in the Tarantula nebula based on a MUSE-NFM mosaic. We aim to deliver the first homogeneous analysis of the most massive stars in the local Universe and to explore the impact of these peculiar objects to the interstellar medium.**

## Resolving the heart of NGC 2070 with MUSE-NFM

The evolution of the Universe is tied to massive stars. They live fast, only a few Myrs, but in very dramatic ways. The energy released during their short lives, but also their deaths in supernova explosions, shapes the chemistry and dynamics of their host galaxies. From the reionisation of the Cosmos through all following epochs of the Universe, massive stars have been mighty sources of ionization. Nonetheless, the stellar evolution of massive O- and B-type stars is far from being well understood, a lack of knowledge that even worsens for the most massive stars (Langer, 2012). These missing pieces in the formation and evolution of massive stars propagate to other fields in astrophysics. Supernova rates, ionization radiation, and chemical yields will be modified according to the evolutionary paths of massive stars. Ultimately, understanding the evolution of star-forming galaxies depends on our ability to, first and foremost, constrain the evolution of massive stars.

Stellar evolution is mainly governed by the initial mass. Nevertheless, other factors can change their paths. Metallicity, rotational velocity, duplicity or strong stellar winds affect their lifetime (Maeder & Meynet, 2000; Langer, 2012). Large systematic surveys are fundamental to unveil the nature of the most massive stars, to constrain the role of these parameters in their evolution, and to provide homogeneous results and landmarks for the theory. Spectroscopic surveys have transformed the field in this direction, yielding large samples for detailed quantitative studies in the Milky Way (e.g. Simon-Diaz et al., 2017) and in the nearby Magellanic Clouds (e.g. Evans et al., 2011). However, massive stars are rare in comparison to smaller stars, and very massive stars (> 70 M☉) are even rarer. The empirical distribution of stars on the upper-part of the Hertzsprung-Russell (HR) diagram remains questionable and more data are essential.

The heart of the Tarantula nebula (NGC 2070) in the Large Magellanic Cloud (LMC) is intrinsically the brightest star-forming region in the Local Group. Its proximity makes it a perfect laboratory to resolve the stellar population and test evolutionary theories. NGC 2070 hosts the most massive stars reported in the literature (Crowther et al., 2010), enclosing the massive cluster R136 in its core. Understandably, NGC 2070 has been of great interest for stellar astrophysicists, and it is considered the Rosetta stone in the field (Schneider et al., 2018). However, in light of the severe stellar crowding in the core of NGC 2070, R136 cluster has largely been omitted in many optical surveys (Evans et al., 2011).

The integral field spectrograph (IFS) MUSE (Bacon et al. 2014) on the VLT has demonstrated to have the capability of resolving crowded stellar fields and taking high quality spectra of thousands of stars in the dense cores of globular cluster (Kamann et al. 2016). In fact, this capability has even been explored out to nearby galaxies (Roth et al. 2020). The large field-of-view and sensitivity of MUSE have allowed us to systematically analyse large stellar populations in unprecedented details, and study the interstellar medium (ISM) (e.g. Weilbacher et al. 2018, Roth et al. 2018). Castro et al. (2018a) presented MUSE wide-field-mode (MUSE-WFM) observations of the central part of NGC 2070. This work provided a homogeneous spectroscopic census of the massive stars in the vicinity of R136. However, the R136 cluster is impossible to resolve with MUSE-WFM.

The MUSE narrow-field-mode (MUSE-NFM) (commissioned in 2018, Leibundgut et al., 2019) has opened a new gate for optical stellar spectroscopy, that until now it was only reachable with the Hubble Space Telescope (HST). The field-of-view of 7.5"x7.5" and an expected spatial resolution close to HST offer unique capabilities to map R136. These capabilities have been successfully tested during ESO program 0104.D-0084. We observed ten fields in NGC 2070 with the MUSE-NFM and mosaicked the R136 cluster in a total of nine pointings. The tenth one was centred on the Wolf-Rayet (WR) system R140. The outstanding performance of the MUSE-NFM and its GALACSI-module in combination with the Adaptive Optics Facility of the VLT provides a spatial resolution that is similar to the HST (~0.07"), however with spectroscopic information for each single pixel, see Fig. 1. A peak in the centre of the cluster reveals how the spatial resolution is even able to resolve the Wolf Rayet (WR ) stars R136a1, 2, and 3, panel b) in Fig. 1.

## R136 cluster: dissecting and modelling

The MUSE-NFM has unveiled a treasure of OB stars for our understanding of the stellar evolution of very massive stars. Based on the integrated light of the MUSE-NFM data cubes, we created a new catalogue for the stellar content of the cluster. The MUSE-NFM catalogue lists approximately 1900 sources in ten fields, with a cut in V band at ~22 mag. A first cross-match with the Hubble Tarantula Treasury Project catalogue (HTTP, Sabbi et al., 2013) showed additional detections and better accuracy for some of the fainter sources close to the brightest stars that are saturated in some of the HST images. The data will allow us to extract the spectra of ~200 stars with good signal-to-noise ratio, enough number and distribution to obtain a clear snapshot of the evolution of OB stars at the age of R136, until approx. $10 M_\odot$ in the HRD.

The MUSE-NFM wavelength range, 4700-9300 Å, does not cover the classic transitions used for spectral classification and stellar atmosphere analysis (Castro et al. 2018a). These canonical features are located at bluer wavelengths than the MUSE cutoff. Nevertheless, MUSE-NFM data offer alternative diagnostics. For O-type and early B-type stars, several He I (4713, 4921, 5876, 6678 Å) and He II (5411, 6683 Å) transitions are included. Hα and Hβ lines and the bluest part of the hydrogen Paschen lines are also visible, offering additional constraints to the effective temperature and gravity.

Previous work (e.g Crowther 2017) shows that stellar analyses with MUSE datasets are possible. Figure 2 displays the analysis of five representative OB stars extracted from one of the central fields in R136. The analysis was performed comparing the observed spectra with a grid of FASTWIND (Puls et al. 2005) synthetic models (see Castro et al. 2018b). The five examples in Fig. 2 show a good match for the key diagnostic lines marked in the plot, i.e. Hβ, HeI 4921 Å and HeII 5411 Å. The residuals observed in the [OIII] 4959, 5007 Å nebular lines are indicative of the difficulty of performing an impeccable sky subtraction, despite the outstanding spatial resolution.

The stellar atmosphere characterization is indeed possible. As shown in Fig. 3, the stars in Fig. 2 match the expected young age of NGC 2070, approx. 2.5Myr. The full analysis of the 200 stars with S/N > 50 will populate the diagram, creating the building blocks to a better understanding of R136 formation and evolution. Only the coolest star (between these five) are displaced from the expected young age, beyond the theoretical main-sequence proposed by Ekström et al. (2012) (see Castro et al 2018b).

## Binary fraction and stellar evolution

Binary stellar evolutionary models have shown drastic effects that the evolution in company can have on each member. Interactions, mass transfer and eventual mergers shape the path and time spent for each star in the HRD (e.g. Wang et al. 2020). If 70% of the OB stars were indeed tied to a companion (Sana et al., 2012), the evolution of massive stars in solitude would be rare. The spectral resolution of MUSE, around 50 km/s, may be considered a limitation before attempting a study of the OB star binary fraction. We expect that massive close/contact spectroscopic binaries have strong radial velocity variations in short epochs, that can be monitored, even with the moderate MUSE spectral resolution.

Our pioneering 0104.D-0084 program was designed to probe the capabilities of the MUSE-NFM in a single epoch. Nevertheless, the observations were spread out in time for technical reasons, so for some of the stars we obtained multiple epochs (see Fig. 1). These overlap regions are priceless to carry out a preliminary test on OB stellar variability. Several resolved spectroscopic binaries were detected in the extracted spectra. Figure 4 shows an example of an O+O binary system, where both He II 5411 Å components are resolved. A Gaussian modelling of both components shows a maximum peak-to-peak variability of ~500 km/s. Close binaries can indeed be characterized at MUSE spectral resolution.

## The carved ISM by the most massive stars

The MUSE-WFM provided new insights in the ISM around the most massive and newly born stars (Castro et al. 2018a). The gas intensity and kinematics was mapped, showing a bi-modal blue and red-shifted motion with respect to the R136 systemic velocity, sketching the ISM in unprecedented detail. A peak in the core revealed red-shifted, possibly infalling material surrounding the strongest X-ray sources (see figure 11 in Castro et al. 2018a). However, the kinematics in the inner part of the cluster could not be explored based on MUSE-WFM spatial resolution.

MUSE-NFM can pierce and dissect the ISM kinematics in the highly dense R136 cluster, where MUSE-WFM capabilities could not push further. Figure 5 shows a coloured image of the central fields using some of the strongest emission lines in MUSE wavelength range: [SII] 6717 Å, Hα and [OIII] 5007 Å . The strong emission in Hα and extended stellar wings of the WR population is clearly visible in Fig. 5, moreover the effect of the radiation carving the ISM at HST-like spatial resolution. We have discovered several new Hα emitters in this first emission map, probably linked to Oe/Be stars or/and pre-main sequence objects. The last ones are expected in an on-going star-forming region such as NGC 2070. New insight in the formation of massive stars and feedback between ISM with strong stellar winds and radiative pressure of the most massive stars will be delivered by MUSE NFM observations.

## More under the hood and future prospects

We are aiming to get a complete snapshot of the stellar evolution of the cluster R136 and to explore the role of different parameters (e.g. duplicity) in the evolution. However, the image quality reached in P104 and the allocated future epochs requested in the forthcoming semesters make us dream further of the possible outcomes. Exploring individual targets of interest in R136 can help us to address open questions, for instance, about the physics driving stellar winds in O-type stars. The rich WR population in R136 will be examined, paying special attention to their strong and extended stellar winds. The data includes the strongest X-rays sources in the field (see Castro et al. 2018a), undoubtedly linked to the most massive WRs: Mk34, R136abc and R140ab (Crowther et al., 2010).

We will explore proper motions in combination with HST data over a baseline of almost ten years since the first HST program (PI: Lennon GO-12499, GO-13359). Mapping the runaway population and their possible links with the cluster will bring insights into the different mechanisms, dynamical ejection and/or binary supernova scenario, that have been suggested to remove the stars from the cluster (e.g. Dorigo Jones et al., 2020).

The MUSE-NFM observations have released the detailed spectroscopic picture of the massive stellar cluster R136 until now. The combined MUSE IFS capabilities (i.e. field-of-view, spatial resolution and spectral coverage) outperform the HST, the only installation capable of resolving the stellar content of R136 at optical wavelengths. This is an outstanding technological achievement, emphasizing the growing role of IFS for stellar astrophysics, whose future is very promising. BlueMUSE (Richard et al., 2019) for the VLT, will open the much desired blue wavelength range such that detailed chemical composition analyses will be possible. The Extremely Large Telescope and next generation of instruments, such as HARMON and MOSAIC, will allow us to leave the Local Group and explore clusters similar to R136 in other galaxies and in even stronger starburst environments.

## Acknowledgements

This project has been funded by the Deutsche Forschungsgemeinschaft (DFG) - CA 2551/1-1. NC and GM gratefuly acknowledge funding from BMBF 05A17BA1. PW and AK from BMBF 05A17BAA for the MUSE-NFM project.

## References

Bacon et al. 2014, The Messenger, 157, 13
Castro et al. 2018a, A&A, 614, 147
Castro et al. 2018b, ApJ, 868, 57
Crowther et al. 2010, MNRAS, 408, 731
Crowther et al. 2017, The Messenger, 170, 40

Dorigo Jones et al., 2020, ApJ, 903, 43
Ekström et al. 2012, A&A, 537, 146
Evans et al., 2011, A&A, 530, 108
Kamann et al., 2016, The Messenger, 164 , 18
Langer, 2012, ARA&A, 50, 107
Langer & Kudritzki, 2014, 2014, 565, 52
Leibundgut et al., 2019, The Messenger, 176, 16
Maeder & Meynet, 2000, ARA&A, 38, 143
Richard et al., 2019, arXiv: 190601657
Roth et al., 2018, A&A, 618, 3
Roth et al., 2020, AN, 340, 989
Sabbi et al., 2013, AJ, 146, 53
Sana et al., 2012, Science, 337, 444
Schneider et al., 2018, Science, 361, 7032
Simon-Diaz et al., 2017, A&A, 597, 22
Puls et al., 2005, A&A, 435, 669
Weilbacher et al., 2018, A&A, 611, 95
Wang et al., 2020, ApJ, 888, L12

## Links

[1] SYCLIST:  https://www.unige.ch/sciences/astro/evolution/en/database/syclist/

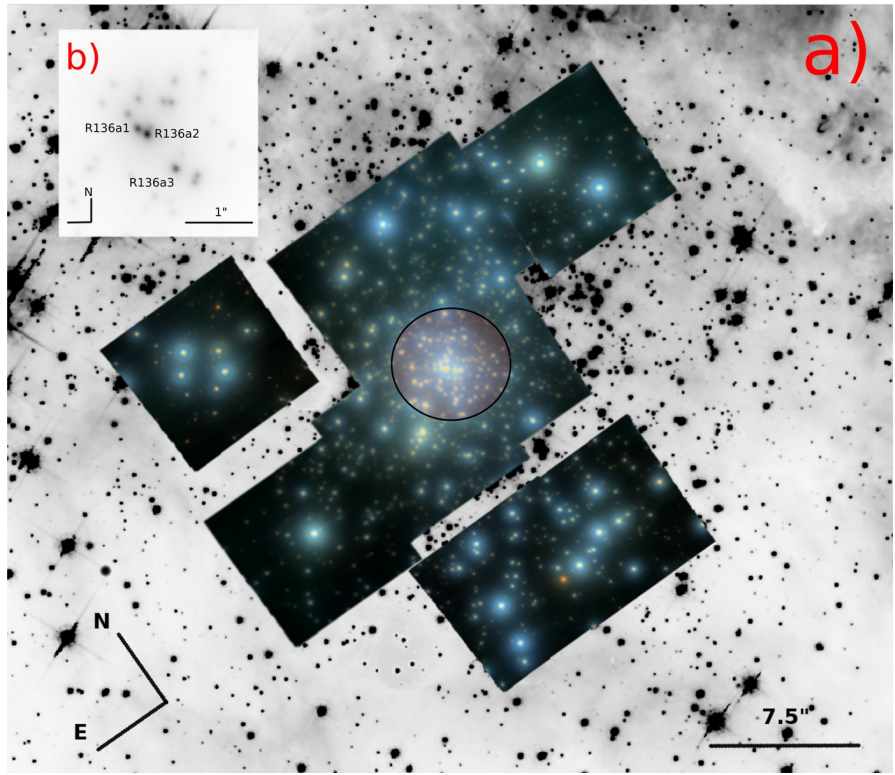

**Figure 1.** Colour-composited mosaic (RGB: I, R and V filters) (a) of nine of the fields observed with MUSE-NFM in the core of NGC 2070. Image quality ranges between 50-80 mas, akin to the HST spatial resolution. The HST image in the F555W band (Sabbi et al. 2013) is displayed in the background. Panel b) zooms into the core of R136 (marked by a circle in mosaic) resolving R136a1, 2 and 3 WR stars.

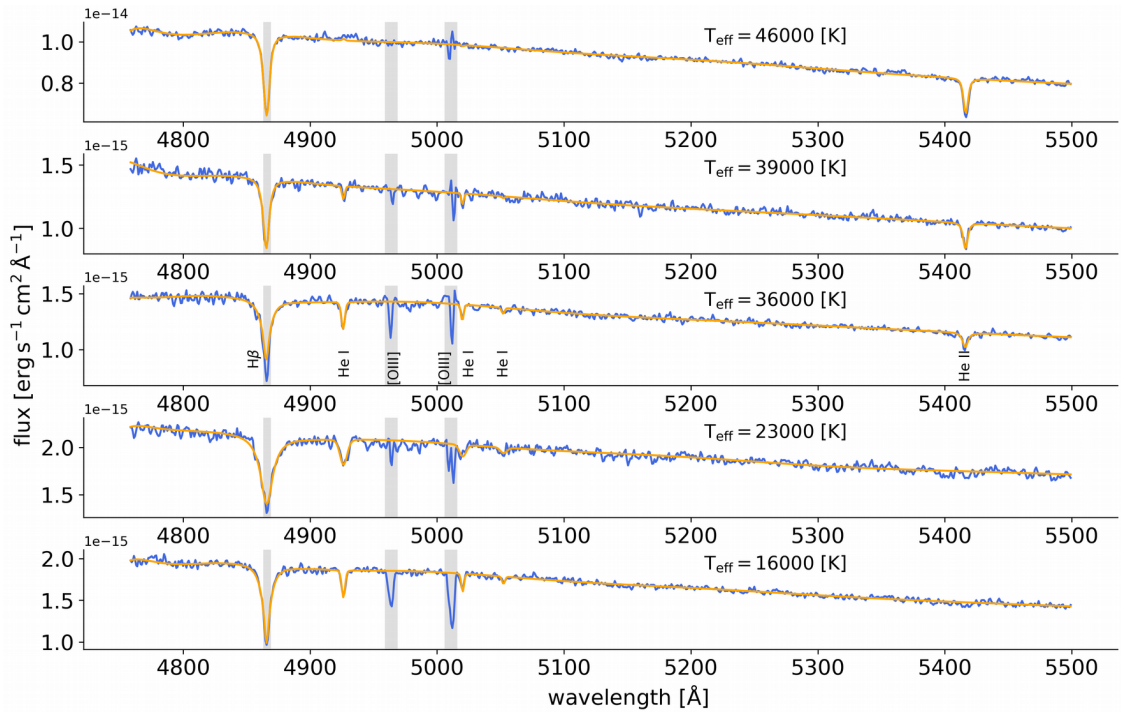

**Figure 2.** OB representative stars extracted from the central field in R136 cluster (blue). The stars were modelled (orange) with a dedicated FASTWIND grid (Puls et al. 2005). The effective temperature and key diagnostic lines are also indicated. The areas that could be affected by the sky subtraction are highlighted (grey shade).

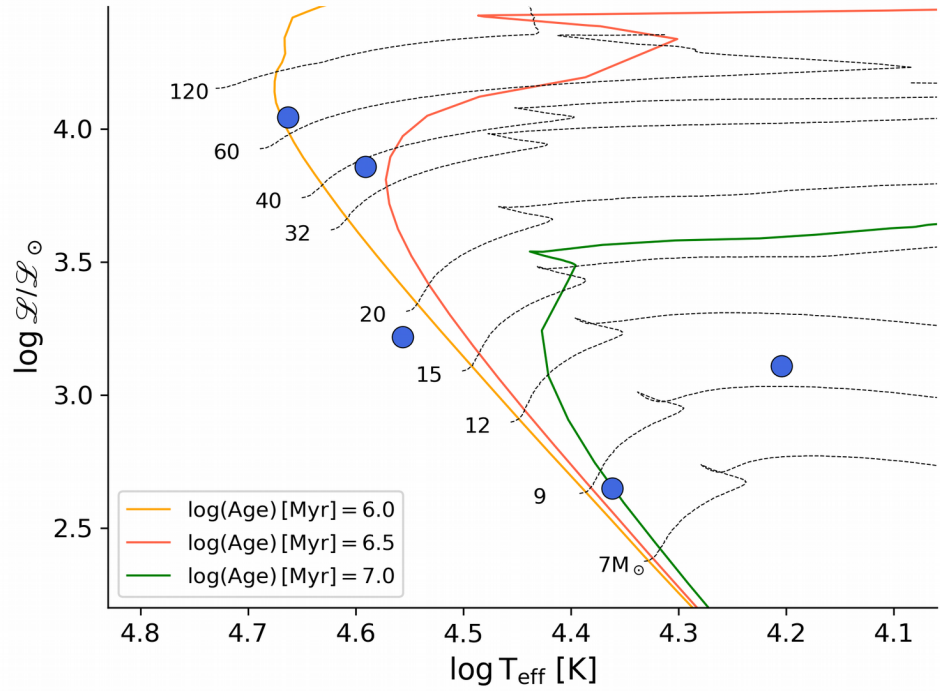

**Figure 3.** Distribution of the representative OB stars (Fig. 2) in the spectroscopic HR diagram (Langer & Kudritzki 2014). The position of the O-type and early B stars mainly match with the expected young age of NGC 2070, approx. 2Myr (e.g. Schneider et al. 2018). Ekström et al. (2012) rotating evolutionary tracks are included in the plot (dotted black lines). The isochrones were calculated using the SYCLIST[1] on-line tool.

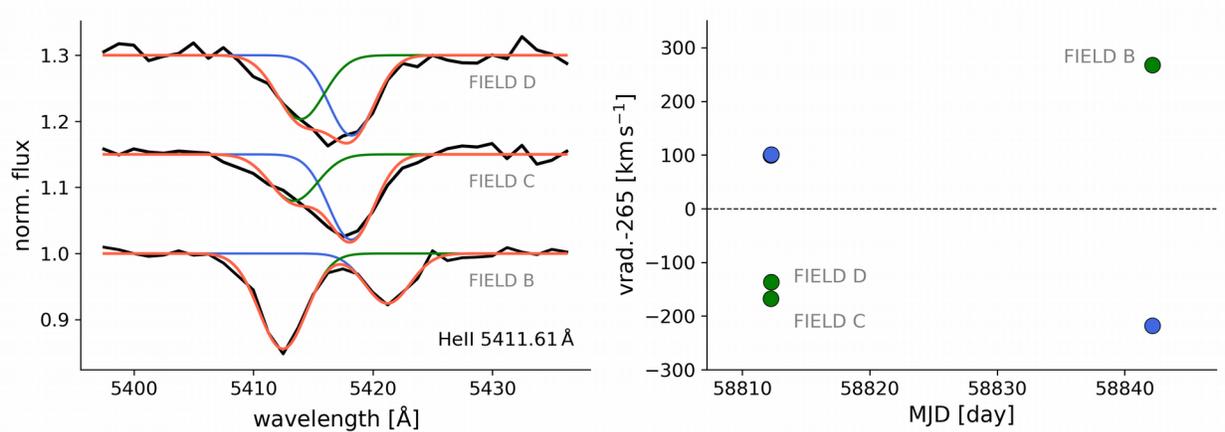

**Figure 4.** O+O spectroscopic binary resolved with MUSE-NFM. On the left, kinematic evolution of HeII 5411 Å (black line) at different epochs in overlapping fields. We perform a preliminary characterization of the binary using Gaussian models (coloured lines). On the right, the velocity of each component is displayed. Note the blue component in field D and C overlaps. The systemic velocity of R136, 265 km/s, was subtracted.

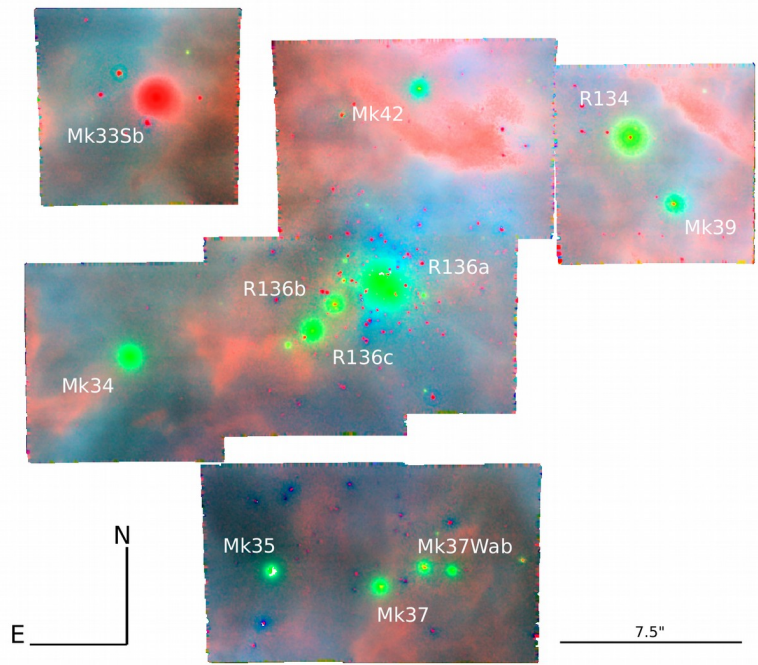

**Figure 5.** MUSE-NFM colour-composited mosaic of the nine fields in the core of NGC 2070 sampling narrow filters around the emission lines: [SII] 6717 Å (red), Hα (green), and [OIII] 5007 Å (blue). The WRs in the field are labelled (e.g. Crowther et al. 2010).